\documentclass[12pt]{article}
\usepackage{graphicx} 
\usepackage{amsmath}
\usepackage[utf8]{inputenc}
\usepackage{braket}
\usepackage{comment}
\usepackage{hyperref}
\usepackage{listings}
\usepackage[a4paper, total={6in, 9in}]{geometry}
\makeatletter
\renewcommand*{\@fnsymbol}[1]{\ifcase#1\or a\or b\or c\or d\or e\or f\or g\or h\or i\or j\else\@ctrerr\fi}
\makeatother

\title{N-Mode Quantized Anharmonic Vibronic Hamiltonians for Matrix Product State Dynamics}

\author{
  Valentin Barandun\thanks{ORCID: 0009-0003-1236-4119}, Nina Glaser\thanks{ORCID: 0000-0002-7477-8099}, and Markus Reiher\thanks{ORCID: 0000-0002-9508-1565} *\\
  \normalsize \textsuperscript{a,b,c}ETH Zürich, Department of Chemistry and Applied Biosciences, \\
  \normalsize Vladimir-Prelog-Weg 2, 8093 Zürich, Switzerland \\
  \normalsize \textsuperscript{b}NNF Quantum Computing Programme,\\
  \normalsize Niels Bohr Institute, University of Copenhagen, Denmark
}

\date{November 5, 2024}

\begin{document}

\maketitle

\begin{center}
{\textbf{Abstract}}
\end{center}
Theoretical predictions of photochemical processes are essential for interpreting and understanding spectral features. Reliable quantum dynamics calculations of vibronic systems require precise modeling of anharmonic effects in the potential energy surfaces and off-diagonal nonadiabatic coupling terms. In this work, we present the $n$-mode quantization of all vibronic Hamiltonian terms comprised of general high-dimensional model representations. 
This results in a second-quantized framework for accurate vibronic calculations employing the density matrix renormalization group algorithm. 
We demonstrate the accuracy and reliability of this approach by calculating the excited state quantum dynamics of maleimide. We analyze convergence and the choice of parameters of the underlying time-dependent density matrix renormalization group algorithm for the $n$-mode vibronic Hamiltonian, demonstrating that it enables accurate calculations of complex photochemical dynamics.

\section{Introduction}
Spectroscopy of vibrationally resolved electronic transitions is a powerful instrument for probing the structure, dynamics, and properties of molecular systems\cite{blanchet1999discerning,doppagne2017vibronic,matselyukh2022decoherence}. Their interpretation necessitates reliable theoretical approaches capable of elucidating and predicting spectral features.
A central challenge in this endeavor is the accurate modeling of the vibronic Hamiltonian that governs optical phenomena, which must capture all relevant interactions and their functional dependencies\cite{bloino2008integrated,baiardi2013general,eng2015spin}. This includes an accurate description of the potential energy surfaces (PES) of the electronic states that enter the vibronic Hamiltonian. Often, approximating the PES as a quadratic function, known as the harmonic approximation, is insufficient to properly describe the vibrational motion of nuclei\cite{mackie2018fully,mccoy2022evidence}.
Additionally, the nonadiabatic coupling terms that appear in the off-diagonal blocks of the vibronic Hamiltonian govern photochemical and photobiological processes\cite{michl1990electronic,schoenlein1991first}. These nonadiabatic coupling terms can be of complex functional form, and therefore, only a non-restrictive approach should allow for an accurate description of these terms. 

Both the anharmonicity in the PES and the functional form of the nonadiabatic coupling terms can be described by the $n$-mode expansion, a many-body expansion allowing the description of the Hamiltonian terms with a high degree of accuracy\cite{christiansen2004second}. Furthermore, it offers a convenient way to formulate second-quantized bosonic Hamiltonians. This second-quantized formulation is a necessity when integrating this approach with some tensor-network methods, which have shown exceptional results in terms of accuracy and computational cost\cite{verstraete2023density}.

One of the most commonly employed tensor-network algorithms is the Density Matrix Renormalization Group (DMRG), which parametrizes the wave function as a tensor-train, also referred to as a matrix product state (MPS), which allows for variational optimizations of bosonic and fermionic wave functions, while scaling polynomially in system size\cite{white1992density,schollwock2011density,baiardi2020density,ma2022density}.

In the MPS parametrization of the wave function, a single tensor is defined per lattice site, which corresponds to a physical degree of freedom such as an orbital or a (vibrational) modal, 
connected by contracting indices. 
The core idea of DMRG is partitioning the optimization problem of diagonalizing the Hamiltonian into a series of many smaller eigenvalue problems. Thereby, a single site of an MPS is optimized at a time while the interaction with the rest of the MPS sites is treated as an effective renormalized basis, whose size is referred to as the bond dimension. 
It is this parameter which determines the accuracy of a DMRG calculation and its computational cost. 

While DMRG provides an efficient framework for obtaining stationary states in large Hilbert spaces, its extension to the time domain, the time-dependent density matrix renormalization group (TD-DMRG), allows for the real- and imaginary propagation of quantum systems within the same tensor network formalism. It enables the direct investigation of non-equilibrium and dynamical phenomena, as well as the calculation of time-dependent spectroscopic quantities such as autocorrelation functions and absorption cross sections. Importantly, TD-DMRG achieves this while efficiently handling large basis sets and complex entanglement structures. This allows for the time-dependent study of molecules with many correlated electrons and vibrational modes\cite{daley2004time,kirino2008time,wolf2015imaginary,ren2018time,baiardi2019large,baiardi2021electron,ren2022time}. Multiple time-dependent variants of the DMRG algorithm have been formulated over the years, such as time-evolving block decimation\cite{vidal2004efficient,white2004real}, adaptive TD-DMRG\cite{feiguin2005time,al2006adaptive,ronca2017time}, the tensor-train split-operator Fourier Transform\cite{greene2017tensor} and the tangent-space formulation of TD-DMRG\cite{baiardi2019large}. The latter is especially suitable for DMRG variants employing MPSs as it exploits their compact structure. The tangent-space formulation of TD-DMRG achieves time evolution based on the Dirac-Frenkel variational principle\cite{moccia1973time,broeckhove1988equivalence}, resulting in projecting the evolved wave function back onto the MPS manifold of a fixed maximum bond dimension\cite{holtz2012alternating,lubich2015time,haegeman2016unifying}.
Since the entanglement entropy tends to increase with time, this procedure introduces truncation errors\cite{de2006entanglement,bardarson2012unbounded}. Therefore, convergence of a TD-DMRG calculation with respect to the maximum bond dimension must be monitored. In this work, we apply the tangent-space TD-DMRG algorithm to the time evolution for a realistic $n$-mode quantized vibronic Hamiltonian. This idea has been previously explored by Shuai and coworkers, while restricting the application of the $n$-mode quantization of anharmonic vibrational potentials to a single ground-state PES and treating off-diagonal terms as constants\cite{ren2021time, wang2021evaluating}. Here, we extend this work to multiple PESs with complex topologies in the vibronic Hamiltonian and allowing complex functional forms of the vibronic coupling terms in $n$-mode quantized form. We demonstrate the applicability of these parametrized Hamiltonians by obtaining accurate vibronic dynamics of molecular systems in combination with a vibronic Hamiltonian, where each term is subject to $n$-mode quantization,
together with TD-DMRG.

\section{Results}

\begin{figure}[htb!]
    \centering
    \includegraphics[width=\linewidth]{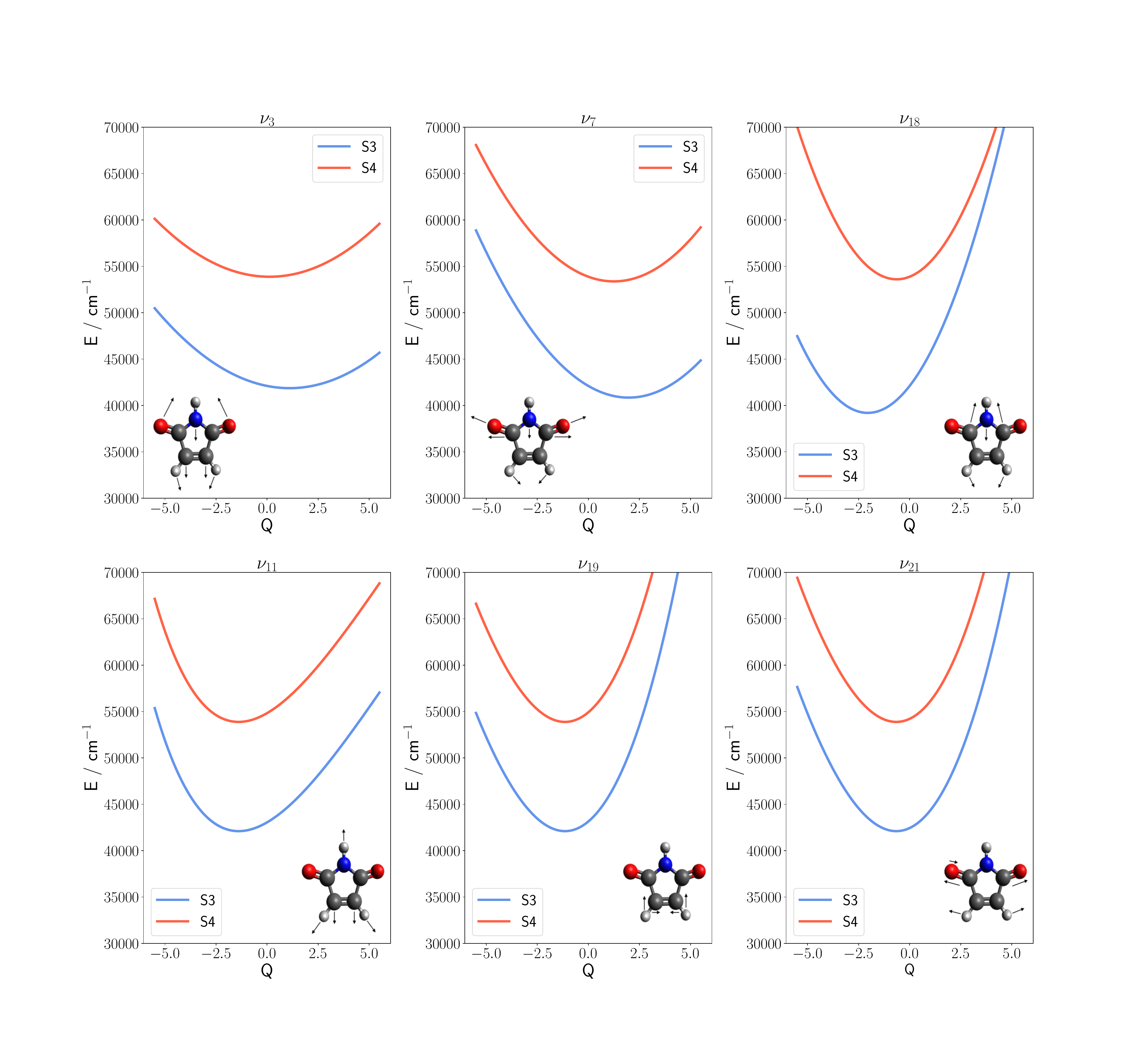}
    \caption{Cuts through the $S_3$ and $S_4$ potential energy surfaces along the selected vibrational mode coordinates. Vibrational mode indices are assigned according to the magnitude of the corresponding harmonic frequencies.  Arrows attached to the molecular structures indicate the displacement of the atoms in each of the selected normal mode coordinate. Atom color code: gray -- carbon, white -- hydrogen, red -- oxygen, blue -- nitrogen.}
    \label{fig:modes}
\end{figure}

To demonstrate the application of the $n$-mode quantization framework for vibronic Hamiltonians with time-dependent tensor-network algorithms, we employ the tangent-space formulation 
of the time-dependent density matrix renormalization group
method \cite{lubich2015time, baiardi2019large} to calculate spectral properties of the maleimide molecule\cite{lubich2015time,haegeman2016unifying, baiardi2019large}.
The $S_0 \rightarrow S_4$ transition was chosen for demonstration purposes of our framework, as it is the absorption band for which experimental spectra are available with good vibrational resolution, which is not the case for the first intense, but very broad $S_0 \rightarrow S_3$ excitation band\cite{seliskar1971electronic}. The initial Franck-Condon excitation $S_0 \rightarrow S_4$ serves as the starting point of the calculation and the subsequent dynamics are are described within the subspace of the $S_3$ and $S_4$ electronic states, which exhibit significant vibronic coupling.
The $n$-mode quantized vibronic Hamiltonian employed for these two electronic excited states includes six out of the total of 24 vibrational modes. These six modes, along which cuts of the PES are depicted in Fig.~\ref{fig:modes}, were selected because they contribute the most to the nonadiabatic coupling terms between the selected electronic states. Hence, these are the essential vibrational modes necessary to reproduce the experimental spectrum. The vibrational modes are numbered in ascending order of the magnitude of their respective harmonic frequencies. The parameters defining the vibronic Hamiltonian were taken from Ref.~\cite{lehr2020role}. As identified in that study, three of the included vibrational modes require anharmonic potentials to be described accurately, making maleimide a suitable system to evaluate our anharmonic formalism.

\begin{figure}[htb!]
    \centering
    \includegraphics[width=.7\linewidth]{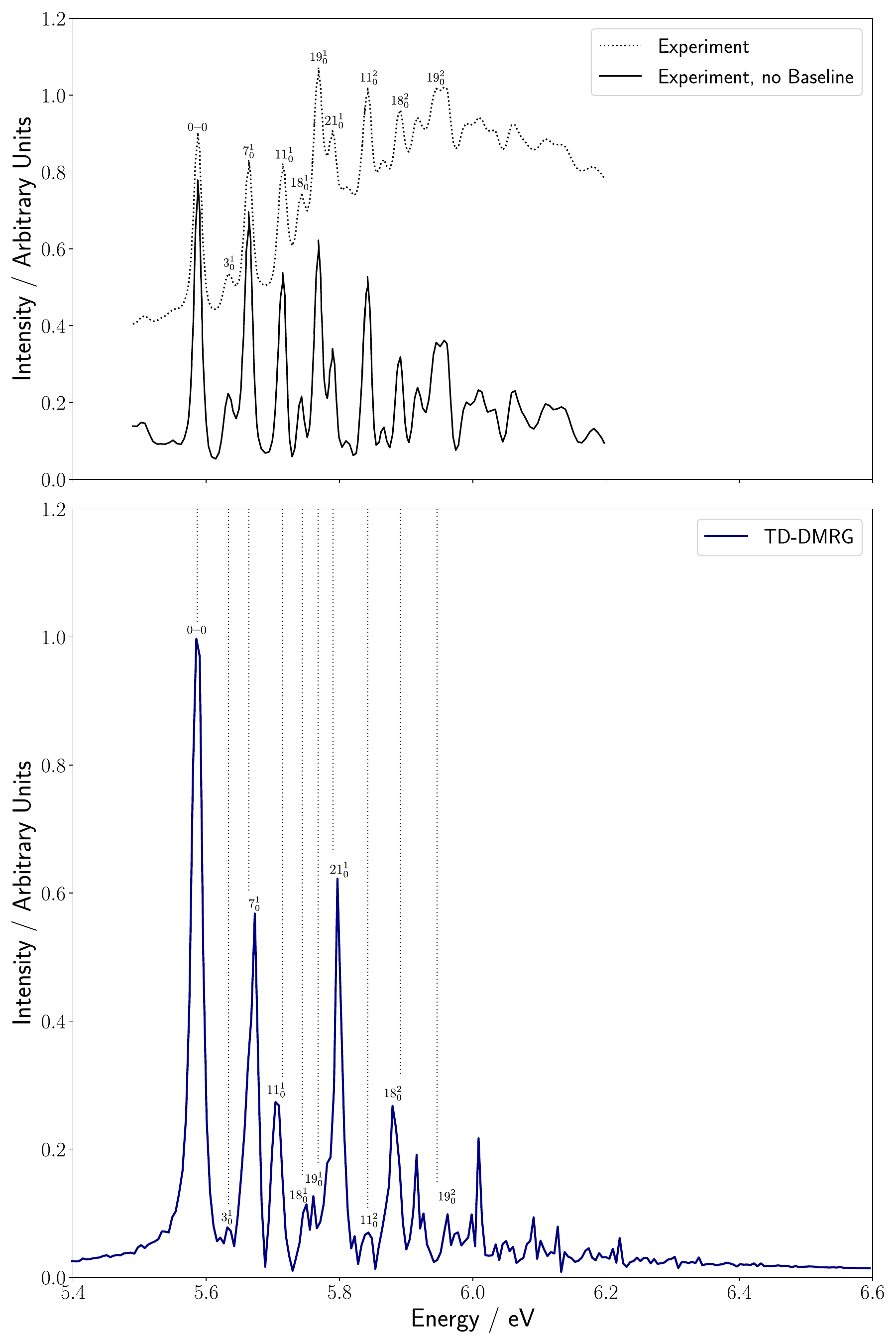}
    \caption{Comparison of the experimental gas-phase absorption spectrum of maleimide taken from Ref.~\cite{lehr2020role} (top) and the TD-DMRG spectrum obtained in this work (with a time step of $0.5$ fs, a total propagation time of $800$ fs, and a bond dimension of 75). Top: Two experimental curves are shown. One represents the raw data and the other shows the spectrum with its envelope subtracted to allow for a better comparison with the calculated spectrum. All fundamental transitions as well as some overtones are labeled according to Ref.~\cite{lehr2020role}. Dotted vertical lines connect the experimental values of these transitions with the calculated results.}
    \label{fig:experiment}
\end{figure}

To assess the reliability of our $n$-mode quantized vibronic framework, we calculated the absorption spectrum of maleimide by Fourier transforming the autocorrelation function obtained by a TD-DMRG calculation of a total propagation time of 800 fs with a maximum bond dimension of 75, shown in Fig.~\ref{fig:experiment}. The calculated spectrum is in good agreement with the experimental results taken from Ref.~\cite{lehr2020role}. Accordingly, this agreement suggests that the essential vibronic couplings and anharmonicities are well described within our framework, allowing for reliable predictions of spectral line shapes and peak positions. We also note that our results are consistent with state-of-the-art multi-layer multi-configurational time-dependent Hartree (ML-MCTDH) 
calculations performed in Ref.~\cite{lehr2020role}.

\begin{figure}[htb!]
    \centering
    \includegraphics[width=.9\linewidth]{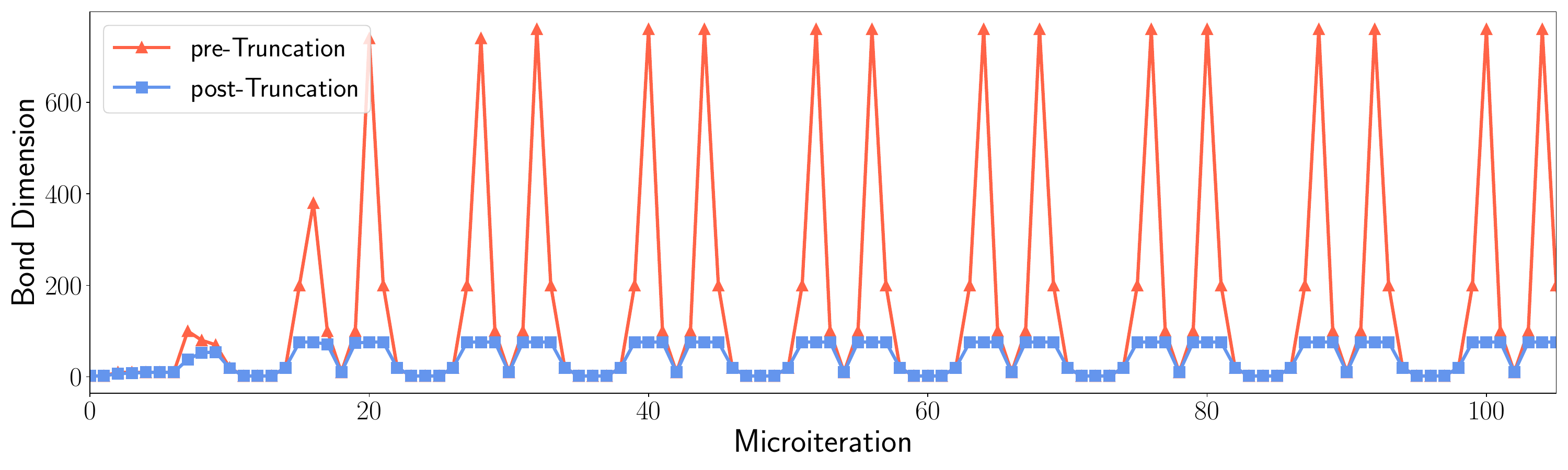}
    \caption{Non-truncated and truncated bond dimensions of the MPS over the first 100 microiterations of the TD-DMRG calculation. The data was obtained by a calculation employing a 0.5 fs time step and a maximum bond dimension of $75$.}
    \label{fig:truncation}
\end{figure}

\begin{figure}[htb!]
    \centering
    \includegraphics[width=.95\linewidth]{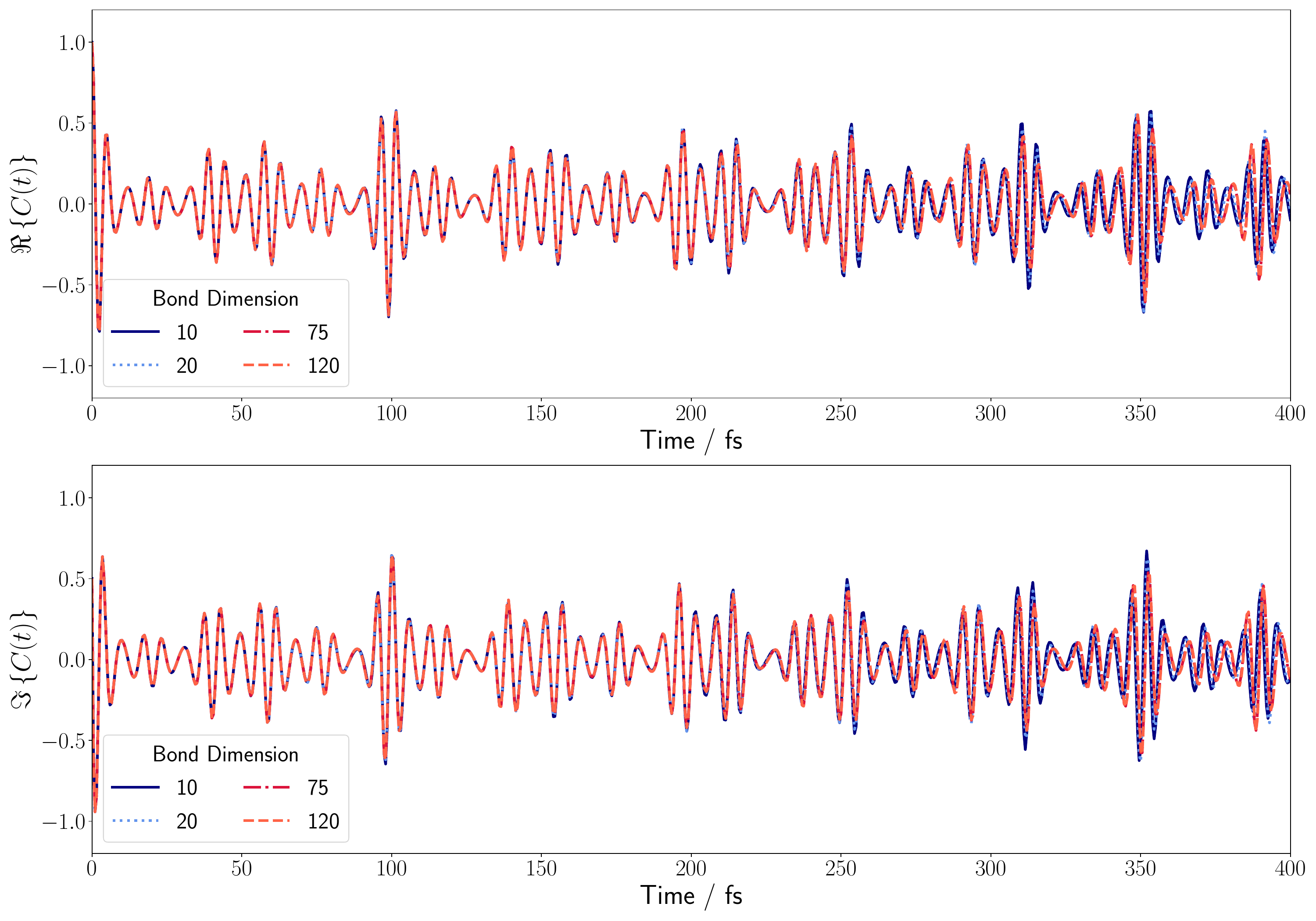}
    \caption{Real and imaginary parts of the autocorrelation function obtained by a TD-DMRG calculation of maleimide upon photoexcitation onto the $S_4$ surface with different values for the maximum bond dimension. The top panel  shows the real and the bottom panel the imaginary part of the autocorrelation function. The results were obtained by employing a time step of $0.5$ fs and propagating the wave function for $400$ fs.}
    \label{fig:auto}
\end{figure}

Since the key convergence parameter of a DMRG calculation is the bond dimension, autocorrelation functions were obtained for different values thereof, as shown in Fig.~\ref{fig:auto}. Although the autocorrelation functions initially coincide, they diverge at longer propagation times. This behavior arises from the fact that the variation of the wave function $\ket{\psi(t)}$ after a time step $\Delta t$ is proportional to $\mathcal{H}\ket{\psi(t)} \times \Delta t$, if $\Delta t \neq 0$. In that case, the exact tensor-train representation of $\mathcal{H}\ket{\psi(t)}$ requires a bond dimension equal to the product of the MPO and MPS bond dimensions. As a result, an accurate and long-time propagation demands increasingly larger MPS bond dimensions if the wave function exhibits considerable entanglement between distant MPS lattice sites. The time evolution of the bond dimension growth and its subsequent truncation by projecting the evolved MPS back onto the manifold spanned by all MPSs of a given bond dimension is illustrated in Fig.~\ref{fig:truncation}. 

\begin{figure}[htb!]
    \centering
    \includegraphics[width=.7\linewidth]{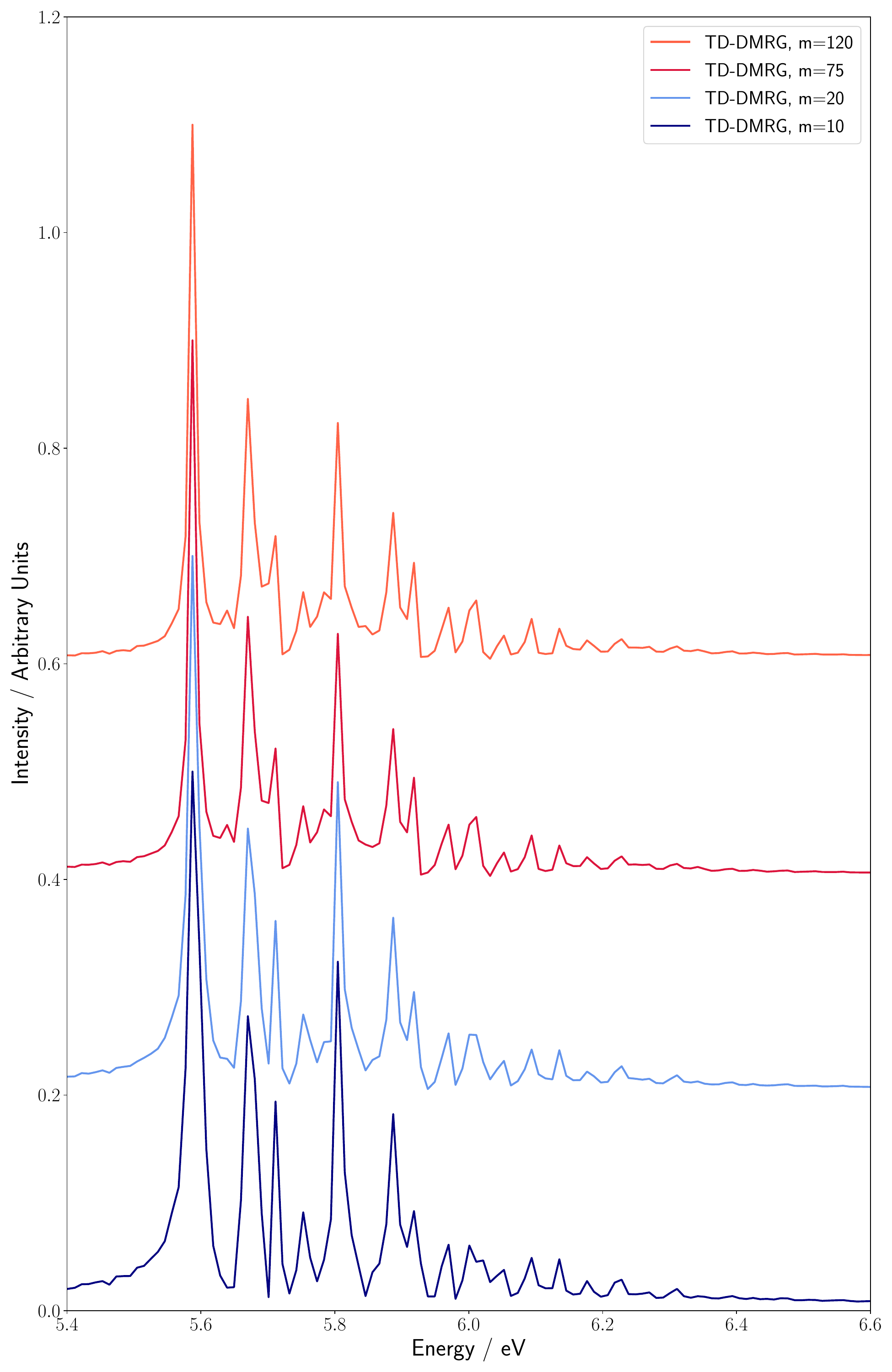}
    \caption{Absorption spectrum of the maleimide molecule upon a Franck-Condon excitation to the $S_4$ electronic surface for different values of the maximum bond dimension m. All calculations were conducted with a time step of 0.5 fs for a total propagation time of $400$ fs.}
    \label{fig:spectrum}
\end{figure}

For the maleimide system studied in this work, comprised of two coupled electronic states and six vibrational degrees of freedom, a maximum bond dimension of $75$ suffices to capture all relevant quantum dynamics. The autocorrelation function has converged for this calculation, exemplified by the near perfect agreement of the autocorrelation function obtained by a calculation with a maximum bond dimension of $75$ and $120$. For shorter time-propagations, smaller bond dimensions are sufficient. For propagations up to $200$ fs, converged calculations can be obtained with a modest bond dimension of $10$, exemplified by the fact that the resulting autocorrelation functions obtained by TD-DMRG calculations with larger bond dimensions match the one obtained with a bond dimension of $10$.
Additionally, in the absorption spectrum obtained by the Fourier transform of the autocorrelation function calculated with a maximum bond dimension of $75$, all fundamental transitions along with the most prominent overtones are present, as illustrated in Fig.~\ref{fig:spectrum}. Performing calculations with bond dimensions significantly lower than that, the $3_0^1$ transition is missing from the calculated absorption spectra. Among the vibrational modes of maleimide, $\nu_3$ exhibits the smallest linear displacement between the $S_0$ and $S_4$ PESs. Since the ground state vibrational wavefunction of the $S_0$ electronic state coincides largely with the node of the first vibrational excited state of the $S_4$ surface, the resulting transition intensity is low in the absence of interstate coupling. Consequently, this results in an inherently weak transition intensity governed by the Franck-Condon overlap between the initial and final states of a given transition. Moreover, this vibrational mode mediates electronic coupling between the $S_3$ and $S_4$ the strongest, giving rise to pronounced entanglement between the electronic MPS sites and the one corresponding to $\nu_3$. Accurately capturing this correlated character requires a sufficiently large bond dimension within the tensor-network representation.

\begin{figure}[htb!]
    \centering
    \includegraphics[width=\linewidth]{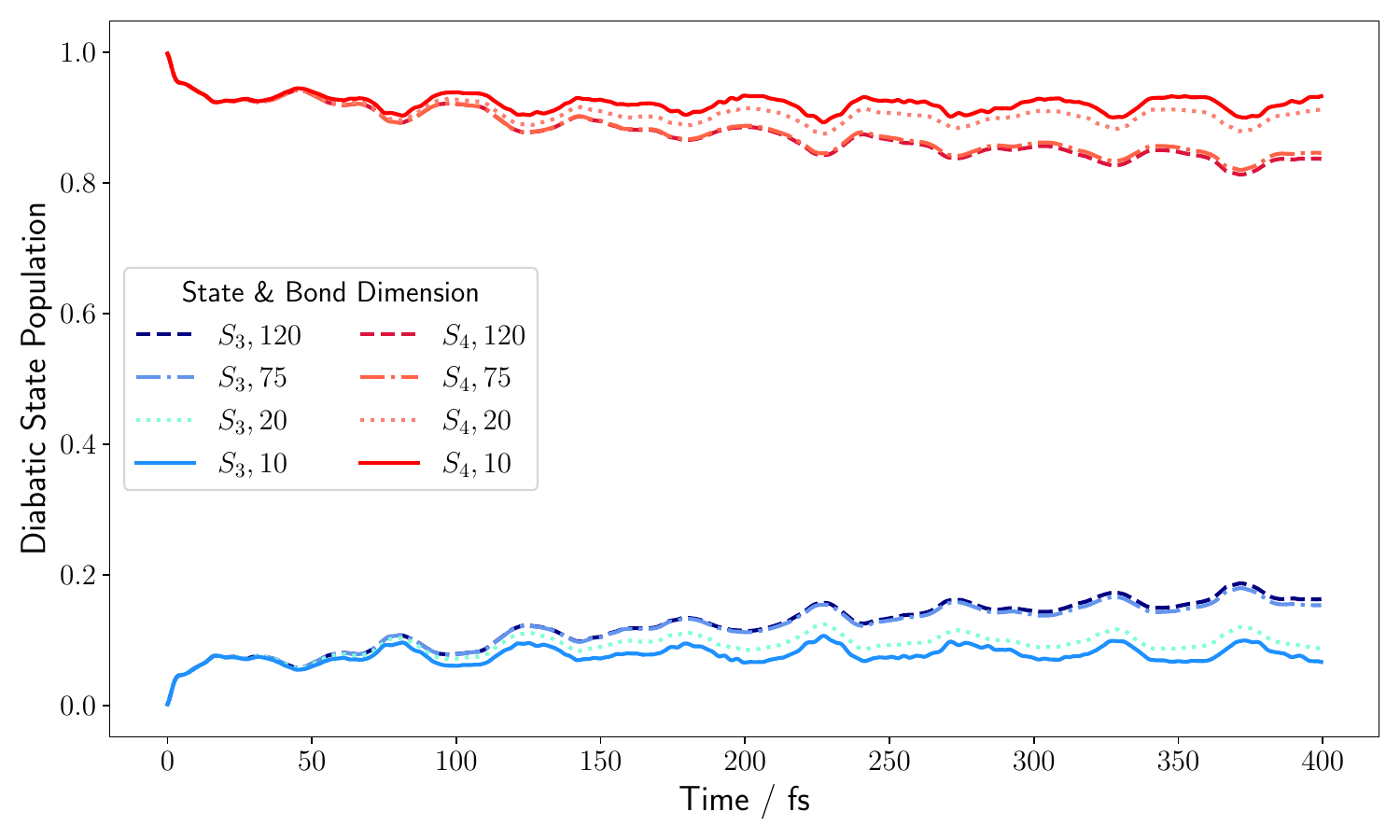}
    \caption{Evolution of the diabatic state population of the maleimide molecule upon photoexcitation onto the $S_4$ electronic state for various values of the maximum bond dimension. The results were obtained for a time step of $0.5$ fs for a total propagation time of $400$ fs.}
    \label{fig:population}
\end{figure}

The diabatic state populations of the $S_3$ and $S_4$ electronic states have been monitored throughout the time-propagation and are shown in Fig.~\ref{fig:population}. The wave packet is initialized on the $S_4$  potential energy surface at $t=0$. Throughout the propagation, a small fraction of the initial population of the $S_4$ state is lost to the $S_3$ state due to non-zero terms in the off-diagonal block in the vibronic Hamiltonian. These terms do not exhibit large magnitudes, explaining the slow population transfer. The population dynamics are consistent with reference calculations obtained with the ML-MCTDH
method\cite{lehr2020role}.

\begin{figure}
    \centering
    \includegraphics[width=\linewidth]{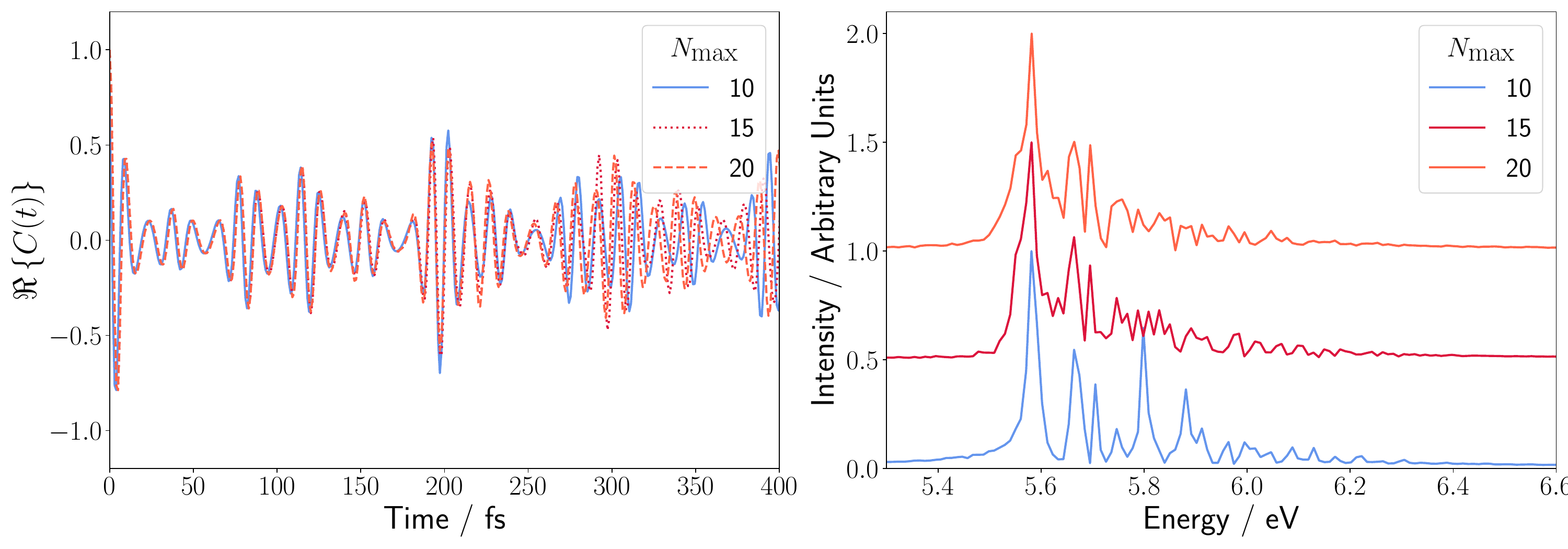}
    \caption{Real part of the autocorrelation function (left) and absorption spectra (right) obtained by a TD-DMRG calculation with a bond dimension of $10$ and a time step of $0.5$ fs for a total propagation time of $400$ fs for different local Hilbert space sizes $N_{\textrm{max}}$.}
    \label{fig:nmax}
\end{figure}

Increasing the bond dimension is not the only route to improve accuracy. Since the integrals corresponding to the diagonal one- and two- body terms of the vibrational Hamiltonians $H_{k_i, h_i}$ and $H_{k_i k_j, h_i h_j}$, as well as the off-diagonal nonadiabatic coupling terms $V_{k_i, h_i}$ and $V_{k_i k_j, h_i h_j}$, defined in Eqs.~\ref{eq:hamilton_nmode} and \ref{eq:coupling_nmode} in the Methods Section, are evaluated numerically, finer grid spacing and larger integration bounds could also enhance accuracy. Moreover, increasing the local Hilbert space dimension $N_{\textrm{max}}$ of each vibrational MPS site reduces the extent of finite basis size errors by including more vibrational basis functions. This effect is illustrated in Fig.~\ref{fig:nmax} , which shows autocorrelation functions and absorption spectra calculated with the same bond dimension but different values of $N_{\textrm{max}}$. Similar to the effect a larger maximum bond dimension has on the quality of the calculated time-dependent quantities, a larger local Hilbert space also enhances the accuracy of a TD-DMRG calculation. The resulting autocorrelation functions initially coincide but diverge at longer propagation times for different values of $N_{\textrm{max}}$. The calculated absorption spectra become more accurate with increasingly large physical basis sizes. This is emphasized by the appearance of the signal corresponding to the $3_0^1$ transition in the calculations with larger local Hilbert spaces. The shape of the $S_3$ potential energy surface poses considerable computational demand due to its large displacement with respect to the ground state minimum energy point, as well as exhibiting the largest frequency correction with respect to the ground state of all the vibrational modes of the $S_4$ PES. Capturing relevant regions of this strongly shifted surface requires a large local basis set. Moreover, extending the local Hilbert space enables the inclusion of higher $S_3$ vibrational excitations that become partially resonant with low-lying $S_4$ vibronic levels, allowing proper mixing and thus a converged description of the transition energy and intensity. This indicates that a sufficiently large $N_{\textrm{max}}$ is essential for accurate long-time calculations of the excited state dynamics of maleimide.

Although increasing $N_{\textrm{max}}$ and the chosen maximum bond dimension improve accuracy, these parameters should be chosen judiciously. In contrast to ground-state DMRG optimizations, which often converge within a few iterations, time-dependent calculations involve many time steps, each corresponding to a sweep, which constitutes the optimization of each MPS lattice site from left to right and back. Consequently, the total computational cost scales with both the bond dimension and the local basis size. It is therefore important to strike a balance between accuracy and computational feasibility by selecting parameters that yield efficient calculations while still capturing all relevant spectral features.

In this work, our framework was applied to the maleimide system to examine its reliability and convergence behavior under well-controlled conditions. However, the methodology is general and can readily be extended to more complex molecular systems. For larger systems with more electronic states and vibrational modes, or in cases with stronger vibronic coupling and MPSs showing larger inter-site entanglement, larger bond dimensions will be necessary to accurately represent the increased complexity of the wave function.

\section{Discussion}
In this work, we applied the $n$-mode quantization framework to a vibronic Hamiltonian of a molecule whose photochemical properties are governed by  multiple anharmonic potential energy surfaces. In principle, this framework also allows the nonadiabatic coupling terms to include complex functional forms. The $n$-mode quantized vibronic Hamiltonian employed with the TD-DMRG algorithm provides an accurate description of vibronic dynamics with significant correlations in the vibrational and vibronic parts of the wavefunction that could not be captured efficiently by harmonic approaches. Conventional vibronic DMRG methods based on canonical quantization employ harmonic oscillator basis functions and represent the PES through a truncated Taylor expansion around a reference molecular structure. Accurately describing strongly anharmonic systems within this canonical harmonic-oscillator based framework requires a very large number of single-particle basis functions, as the harmonic expansion poorly approximates regions far from the reference point. Consequently, the solution of the Schr\"odinger equation can become computationally intractable even for moderately sized molecules that exhibit pronounced anharmonicities. Moreover, since a Taylor expansion is intrinsically local, complex potential energy surfaces, such as double-well or multi-minimum surfaces, cannot be reliably represented. 

Expressing the vibronic wavefunction as a tensor network allows us to mitigate the curse of dimensionality that limits quantum mechanical many-body methods. Applying our approach to the photo-induced dynamics of maleimide allowed us to study how the bond dimension $m$ and the number of single particle basis functions per mode $N_{\textrm{max}}$, which both govern the accuracy and cost of a vibronic DMRG calculation, should be chosen to achieve convergence for moderately sized molecules.
Other state-of-the-art methods for vibronic dynamics include ML-MCTDH\cite{meyer1990multi,manthe1992wave, meyer2003quantum, wang2003multilayer, manthe2008multilayer} and surface-hopping approaches\cite{tully1990molecular,drukker1999basics,tully2012perspective}.
In contrast to surface hopping, our method offers a fully quantum-mechanical description of the coupled electronic and nuclear degrees of freedom. Compared to the reference approach ML-MCTDH, the $n$-mode quantized vibronic Hamiltonian in combination with TD-DMRG provides more rigorous and systematic error control through tuning of the bond dimension and by monitoring the discarded singular values that arise during the truncation steps intrinsic to DMRG. While tensor-network methods have opened the door to treating larger anharmonic vibrational and vibronic systems, increasing system size remains a significant challenge. Further improvements can be achieved by applying concepts from quantum information theory to optimize the mapping and ordering of vibrational modes and electronic basis functions on the DMRG lattice. 
Such analyses have already been demonstrated to be very useful for purely electronic\cite{boguslawski2012entanglement, boguslawski2013orbital, stein2016automated} and purely vibrational Hamiltonians\cite{glaser2024vibrational} and will serve as a foundation for extending the accessible system size in vibronic calculations. In addition, finite-temperature DMRG algorithms can be incorporated with minimal modifications to the present framework\cite{verstraete2004matrix, feiguin2005finite}, enabling the simulation of temperature-dependent spectroscopic features relevant to actual experimental conditions.

\section{Methods}
\label{sec:methods}

\subsection{$n$-Mode Quantized Vibronic Hamiltonian}

Vibronic dynamics involving multiple electronic states and vibrational degrees of freedom can be described with a general vibronic Hamiltonian of the following form
\begin{equation}
  \mathcal{H}_\text{vibronic} =
  \begin{bmatrix}
    \mathcal{H}_1(\mathbf{Q})
    & \mathcal{V}_{12}(\mathbf{Q})
    & \cdots
    & \mathcal{V}_{1 N_\text{el}}(\mathbf{Q}) \\
    \mathcal{V}_{21}(\mathbf{Q})
    & \mathcal{H}_2(\mathbf{Q})
    &
    & \\
    \vdots
    &
    & \ddots
    & \\
    \mathcal{V}_{N_\text{el} 1}(\mathbf{Q})
    &
    &
    & \mathcal{H}_{N_\text{el}}(\mathbf{Q}) \\
  \end{bmatrix},
  \label{eq:VibronicHamiltonian}
\end{equation}
where $\mathcal{H}_{\alpha}(\mathbf{Q}) = \mathcal{T}_{\alpha}(\mathbf{Q}) + v_{\alpha}(\mathbf{Q})$ is the vibrational Hamiltonian associated with the $\alpha$-th electronic state (with a kinetic and a potential energy term, $\mathcal{T}_{\alpha}(\mathbf{Q})$ and $v_{\alpha}(\mathbf{Q})$, respectively). $\mathcal{V}_{\alpha\beta}(\mathbf{Q})$ is the nonadiabatic coupling between the electronic states $\alpha$ and $\beta$. Indices $\alpha$ and $\beta$ range from 1 to the number of electronic states $N_{\textrm{el}}$. $\mathbf{Q} = \{Q_1, Q_2, \cdots,Q_M\}$ denotes the set of $M$ vibrational degrees of freedom of the system. Various approximations are made in such a model Hamiltonian, which enter through the specific definition of the vibrational Hamiltonians on the diagonal blocks in Eq. ~\eqref{eq:VibronicHamiltonian} and the nonadiabatic coupling terms on the off-diagonal blocks.

In a vibronic DMRG calculation, a second-quantized framework that allows for the implementation of arbitrary functional forms of the potential energy surfaces and nonadiabatic coupling terms is crucial. The $n$-mode quantization scheme offers a suitable approach for this purpose\cite{christiansen2004second}. The potential energy surfaces and the nonadiabatic coupling terms are expressed in a high-dimensional model representation with the degrees of freedom corresponding to the vibrational modes of the system. The expansion is written as a sum over grouped terms, categorized by the number of degrees of freedom each term depends on. The $n$-mode expansion of an arbitrary function $\mathcal{F}(\mathbf{Q})$ depending on $M$ degrees of freedom $\mathbf{Q}$ is given by
\cite{christiansen2004second} 
\begin{equation}
    \mathcal{F}(\mathbf{Q}) = \sum_i^M \mathcal{F}_1^{[i]}(Q_i) + \sum_{i<j}^M \mathcal{F}_{2}^{[ij]}(Q_i,Q_j) + \sum_{i<j<k}^M \mathcal{F}_3^{[ijk]}(Q_i, Q_j, Q_k) \cdots
\end{equation}
$\mathcal{F}_1^{[i]}(Q_i)$ depends on one internal coordinate and accounts for the variation of the function to be approximated with respect to that coordinate. The term $\mathcal{F}_2^{[ij]}(Q_i,Q_j)$ depends on two internal coordinates and accounts for the variation of the function with respect to a simultaneous change of the coordinates $Q_i$ and $Q_j$. Higher order terms follow analogously. The vibrational Hamiltonian of an arbitrary electronic state and a nonadiabatic coupling term between two electronic states, can be expressed in second-quantized form in $n$-mode quantization as 
\begin{equation}\label{eq:hamilton_nmode}
  \mathcal{H}_{\text{nmode}} =  \sum_{i=1}^M \sum_{k_i,h_i=1}^{N_i} H_{k_i, h_i}^{[i]}
  \hat{b}_{k_i}^\dagger \hat{b}_{h_i}
  + \sum_{i=1}^M\sum_{i<j}^{M} \sum_{k_i,h_i=1}^{N_i} \; \sum_{k_j,h_j=1}^{N_j} H_{k_i k_j, h_i h_j}^{[i,j]}
  \hat{b}_{k_i}^\dagger \hat{b}_{k_j}^\dagger \hat{b}_{h_i} \hat{b}_{h_j} + \ldots \, ,
\end{equation}
and
\begin{equation}\label{eq:coupling_nmode}
  \mathcal{V}_{\text{nmode}} =  \sum_{i=1}^M \sum_{k_i,h_i=1}^{N_i} V_{k_i, h_i}^{[i]}
  \hat{b}_{k_i}^\dagger \hat{b}_{h_i}
  + \sum_{i=1}^M\sum_{i<j}^{M} \sum_{k_i,h_i=1}^{N_i} \; \sum_{k_j,h_j=1}^{N_j} V_{k_i k_j, h_i h_j}^{[i,j]}
  \hat{b}_{k_i}^\dagger \hat{b}_{k_j}^\dagger \hat{b}_{h_i} \hat{b}_{h_j} + \ldots \,
\end{equation}
respectively.
Indices $i$ and $j$ run over the $M$ vibrational modes and indices $k_i$, $h_i$, $k_j$ and $h_j$ run over the number of single particle basis functions for vibrational mode $i$ and $j$, respectively. The bosonic creation and annihilation operators are defined in Ref.~\cite{glaser2023flexible}.
The one- and two-body integrals, in the case of a real-valued basis set, are defined as 
\begin{equation}
    H_{k_i,h_i}^{[i]}=\int_{-\infty}^{+\infty}\phi_i^{k_i}(Q_i)(\mathcal{T}(Q_i)+v_1^{[i]}(Q_i))\phi_i^{h_i}(Q_i)\mathrm{~d}Q_i
\end{equation}

\begin{equation}
    \begin{aligned}H_{k_ik_j,h_ih_j}^{[i,j]}&=\int_{-\infty}^{+\infty}\int_{-\infty}^{+\infty}\phi_i^{k_i}(Q_i)\phi_j^{k_j}(Q_j)v_2^{[i,j]}(Q_i,Q_j)\phi_i^{h_i}(Q_i)\phi_j^{h_j}(Q_j)\mathrm{~d}Q_i\mathrm{~d}Q_j\end{aligned}
\end{equation}

\begin{equation}
    V_{k_i,h_i}^{[i]}=\int_{-\infty}^{+\infty}\phi_i^{k_i}(Q_i)\mathcal{V}^{[i]}_1(Q_i)\phi_i^{h_i}(Q_i)\mathrm{~d}Q_i
\end{equation}

\begin{equation}
    \begin{aligned}\mathcal{V}_{k_ik_j,h_ih_j}^{[i,j]}&=\int_{-\infty}^{+\infty}\int_{-\infty}^{+\infty}\phi_i^{k_i}(Q_i)\phi_j^{k_j}(Q_j)\mathcal{V}_2^{[i,j]}(Q_i,Q_j)\phi_i^{h_i}(Q_i)\phi_j^{h_j}(Q_j)\mathrm{~d}Q_i\mathrm{~d}Q_j,\end{aligned}
\end{equation}
where $\mathcal{T}(Q_i)$ is the kinetic energy term of a vibrational mode $Q_i$, $v_1^{[i]}$ and $v_2^{[i,j]}$ are the one- and two-body terms 
of the $n$-mode expansion of the potential energy surfaces, while $\mathcal{V}^{[i]}_1(Q_i)$ and $\mathcal{V}_2^{[i,j]}(Q_i,Q_j)$ are the one- and two body terms of the $n$-mode expansion of the nonadiabatic coupling terms. Higher order terms follow analogously.
These integrals can be evaluated with an arbitrary set of single-particle basis functions $\{\mathbf{\phi}_i^{k_i}\}$, emphasizing the advantage of the second-quantized $n$-mode Hamiltonian, for which a suitable basis set can be chosen for a given problem. Since the $n$-mode expansion of a function rapidly converges for many systems with an appropriate choice of coordinates and functions,\cite{rabitz1999general,alics2001efficient,manzhos2006random} the $n$-mode quantized vibronic Hamiltonian in second quantization enables the efficient numerical description of high-dimensional vibronic problems. In particular, its compatibility with the second-quantized MPS/MPO formulation of DMRG makes it a powerful tool for exploring the time evolution and spectral properties of complex molecular systems.

\subsection{Vibronic Matrix Product State}

\begin{figure}[htb!]
    \centering
    \includegraphics[trim=0cm 5cm 0cm 6cm, clip, width=0.8\textwidth]{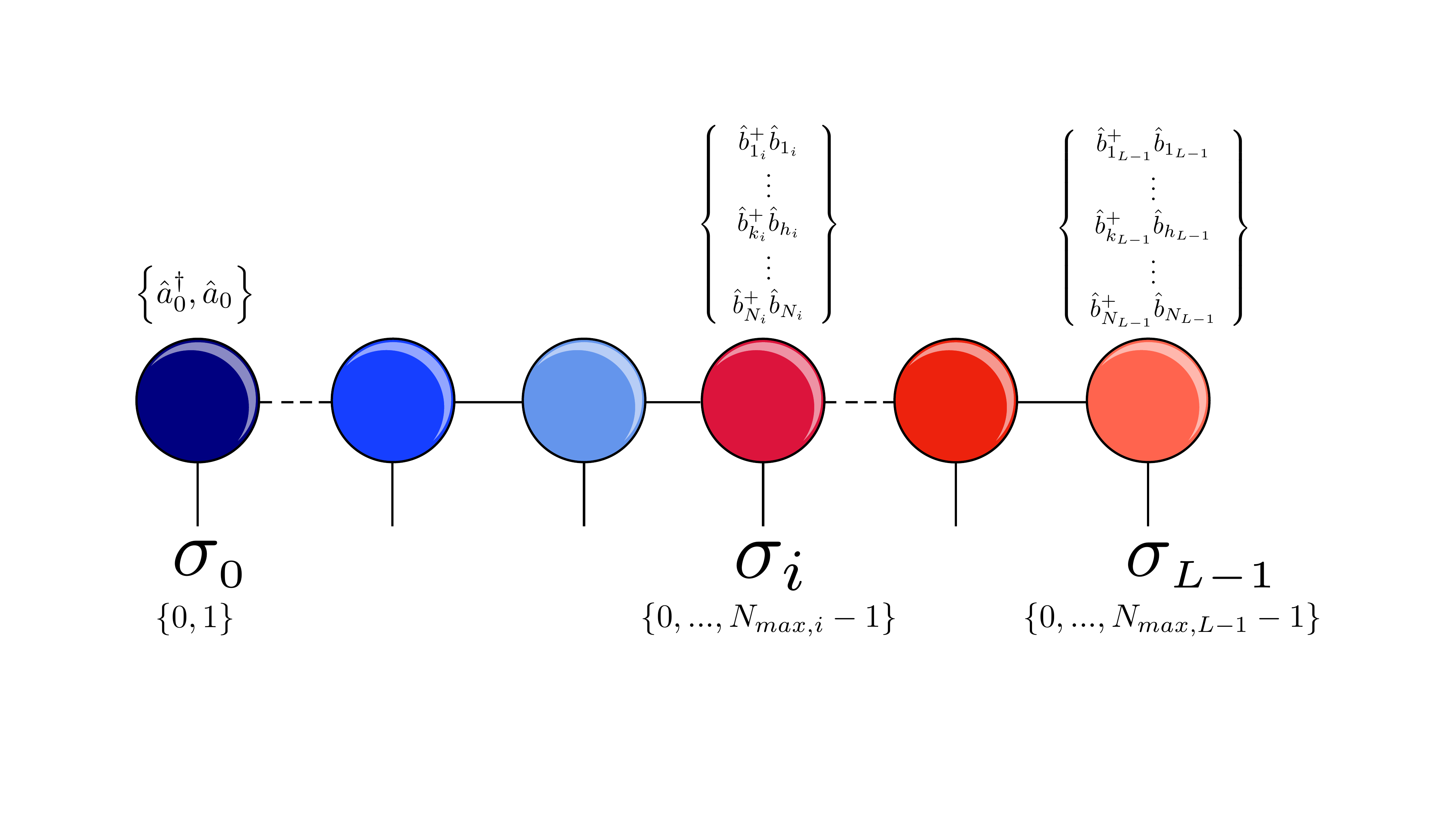}
    \caption{Graphical representation of the vibronic MPS. The first sites are electronic states (color-coded in red) followed by the vibrational sites (color-coded in blue). The electronic sites can have occupations $0$ or $1$ encoding whether the electronic state is populated or not. The quantum numbers of the physical basis $\sigma$ of the vibrations can be any integer from $0$ to $N_{\mathrm{max}} - 1$, where $N_\mathrm{max}$ denotes the number of vibrational basis functions per mode. The corresponding operators acting on each site are indicated above the individual MPS sites.}
    \label{fig:mps}
\end{figure}

The vibronic MPS was constructed as illustrated in Fig.~\ref{fig:mps}: The $N_{\textrm{el}}$ electronic sites, which each correspond to an electronic state, placed at the beginning of the DMRG lattice, followed by the $M$ bosonic sites representing the vibrational modes of the system. The operators applied to the MPS sites are the electronic creation and annihilation operators $\hat{a}_\gamma^{\dagger}$, $\hat{a}_\gamma$ for the $\gamma$-th electronic state and the bosonic creation and annihilation operators $\hat{b}_{k_i}^\dagger$, $\hat{b}_{k_i}$, which create and destroy occupations in the $k_i$-th vibrational basis function of the vibrational site corresponding to mode $i$. In a naive implementation this would yield operators of dimension $d=(N_i + 1) \times (N_i + 1)$, where $N_i$ is the number of vibrational basis functions describing a single vibrational mode $i$. Since an auxiliary vacuum state is needed to describe a depopulation from basis function $k_i$ to the vacuum or a population of a basis state $k_i$ from the vacuum, the dimension of the matrices responsible for these operations have to be $N_i + 1$, rather than $N_i$. By not registering the single bosonic operators separately, but the operator product of the form $\hat{b}_{k_i}^{\dagger}\hat{b}_{h_i}$ as a composite operator (as it appears in Eqs.~\eqref{eq:hamilton_nmode} and \eqref{eq:coupling_nmode}) renders the vacuum state redundant, resulting in product operators of dimension $d = N_i \times N_i$. The vibronic MPS is of U1 symmetry since there is electronic particle conservation.

\subsection{Measurements}

The absorption spectra, assuming a constant dipole moment, were obtained by Fourier transforming the autocorrelation function, 
\begin{equation}
I(\omega) =  \int_0^{\infty} e^{i\omega t} C(t) \, dt =  \int_0^{\infty} e^{i\omega t} \langle \psi(0) | \psi(t) \rangle \, dt
\end{equation}
and subsequently plotting the absolute magnitude of the complex-valued quantity $I(\omega)$ after shifting the $0$--$0$ transition to match the experimental excitation energy. Further post-processing of the Fourier-transformed function, such as zero-padding or the convolution with a broadening function, was not necessary since this was not required to obtain results matching the experimental spectra. The population of the diabatic electronic states were measured by evaluating the expectation value of the electronic number operator
\begin{equation}
    \mathcal{\hat{N}}_{\gamma} = \hat{a}_\gamma^{\dagger}\hat{a}_\gamma
\end{equation}
where $\hat{a}_\gamma^{\dagger}$ and $\hat{a}_\gamma$ are the creation and annihilation operators, respectively, acting on the electronic MPS site $\gamma$. Essentially, this projects the electronic part of the vibronic wave function onto the electronic state $\gamma$, yielding in the probability of encountering the system in the diabatic electronic state $\gamma$.

\subsection{Computational Details}

All quantum dynamics calculations were carried out with the DMRG software package \textsc{QCMaquis}\cite{szenes2025qcmaquis}. The time step for the time evolution should generally be chosen such that it captures the fastest relevant oscillations in the system under study. In our case, a time step of $0.5$ fs delivered accurate results. A total number of $800$ sweeps (a sweep consitutes an optimization of each lattice site from the beginning of the lattice to its terminus and back) 
was employed in most calculations (unless stated otherwise), resulting in a total propagation time of $400$ fs. For all calculations, the two-site DMRG integrator was employed. The integrals that appear in the definition of the matrix product operator representation of the Hamiltonian were evaluated numerically. For these integrations, the basis functions were chosen to be the vibrational wave functions of the vibrational modes of the ground electronic state. 

\section*{Acknowledgments}
M.R. and V.B. have been financially supported by the Swiss National Science Foundation (grant no. 200021\_219616).
N.G. gratefully acknowledges support by the Novo Nordisk Foundation, Grant number NNF22SA0081175, NNF Quantum Computing Programme.



\end{document}